\documentclass[journal]{IEEEtran}
\ifCLASSINFOpdf
\else
\fi
\usepackage{amssymb}
\usepackage{amsmath}

\usepackage{latexsym,color,amsfonts}
\usepackage{pgfplots,graphicx}
\usetikzlibrary{shapes.geometric,calc}
\usetikzlibrary{spy}

\usepackage{scalefnt}
\usepackage{subcaption}

\usepackage{todonotes}

\renewcommand{\vec}[1]{{\boldsymbol#1}}
\newcommand{\ie}{\textit{i.e.}\/, }
\newcommand{\eg}{\textit{e.g.}\/, }

\begin{document}
%
\title{Evaluation of the Electric Polarizability for Planar
Frequency Selective Arrays}
%
%
%

\author{Andrei~Ludvig-Osipov*,~\IEEEmembership{}
        and~B.L.G.~Jonsson,~\IEEEmembership{}
\thanks{A. Ludvig-Osipov and B.L.G. Jonsson are with the School of Electrical Engineering and Computer Science, KTH Royal Institute of Technology, SE10044 Stockholm,
Sweden e-mail: osipov@kth.se.}
\thanks{}}

%
%

\markboth{Electric Polarizability of Planar Arrays}
{Shell \MakeLowercase{\textit{et al.}}: Bare Demo of IEEEtran.cls for IEEE Journals}
%



\maketitle

\begin{abstract}
This paper presents a method to evaluate the static electric polarizability of two-dimensional infinitely periodic metal patch arrays with dielectric substrate. Static polarizabilities are used in several design applications for periodic structures such as estimation of the bandwidth limitations for frequency-selective structures or prediction of the radiation properties for antenna arrays. The main features of the proposed method are its numerical efficiency and a deep insight into the physics of the fields interacting with the structure. 
We provide derivation and analysis of the method, and its verification against two another commercial solver-based approaches for various structure geometries.
Additionally, we suggest the guidelines for applying the method to bandwidth optimization of frequency selective structures and illustrate this with an example.

\end{abstract}

\begin{IEEEkeywords}
Polarizability, frequency selective surfaces, scattering, periodic structures, sum rules.
\end{IEEEkeywords}

%
\IEEEpeerreviewmaketitle

\section{Introduction}
%
%
%
%
Planar frequency selective structures appear in a number of electromagnetic devices, such as filters~\cite{dickie2011}, absorbers~\cite{munk}, polarizers~\cite{ericsson2014}, beam splitters and reflection/transmission arrays~\cite{mittra}.
They are widely used in antennas as components of reflectors, radomes and as antenna arrays themselves~\cite{munk2}.
High impedance surfaces and several metamaterials are implemented as periodic structures~\cite{liu2009}.
Furthermore, the phenomenon of extraordinary transmission through periodic structures has attracted a renewed scientific interest~\cite{gustafsson2011}.
The fundamental principles of frequency selective surfaces are related to optical diffraction gratings, discovered in the late 18th century~\cite{rittenhouse}.
Since then, frequency selective surfaces for optical and radio-frequency spectra has been thoroughly studied and described in the literature, see e.g. ~\cite{munk,mittra}.

Limitations of the bandwidth for periodic structures has begun to be investigated~\cite{gustafsson2011,bernland2011,sjoberg2009} in terms of sum rules.
The bounds connect dynamic behavior of the structure to its static electric and magnetic polarizabilities.
To efficiently and accurately determine polarizabilities is thus essential in establishing the bounds~\cite{sjoberg2008,jelinek}.

In this paper we propose a computationally effective method to estimate the electric polarizability of a frequency selective structure consisting of an array of patches.
The distinct properties of the proposed method are its numerical efficiency and analyticity, providing insight in the physics of the frequency selective stuctures. This is beneficial for optimization of such structures.


\section{Problem Description}

\begin{figure}[h]
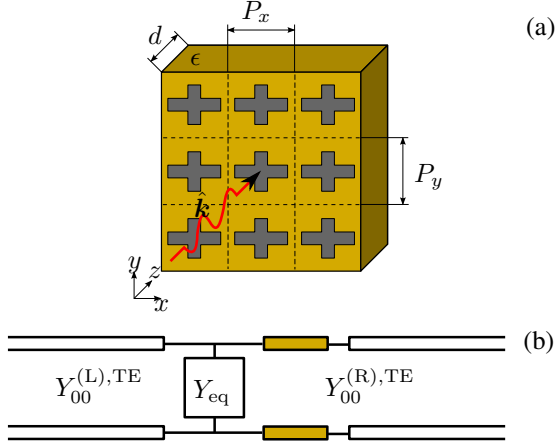

  \centering
  \begin{minipage}{0.45\textwidth}
  \centering
  \begingroup
	\tikzset{every picture/.style={scale=0.9}}
	\input{struct.tikz}
  \endgroup
  \vspace{-15pt}
  \centering
  \begingroup
	\tikzset{every picture/.style={scale=1.5}}
	\input{eq_scheme.tikz}
  \endgroup
  \end{minipage}
  \vspace{15pt}
  \caption{(a) A periodic array of patches on a dielectric substrate of thickness $d$ and permittivity $\epsilon$; the unit cell size is $P_x\times P_y$, $\hat{\vec{k}}$ is a wavevector of an impinging field.
  			(b) An equivalent scheme for an impinging mode.
}
  \label{fig:R_vs_freq}
\end{figure}

The goal of this paper is to develop an efficient method to determine the electric polarizability for a planar two-dimensional array.
We consider a negligibly thin layer of PEC patches, placed parallel to the $xy$-plane with a rectangular unit cell of size $P_x\times P_y$ and a dielectric substrate thickness $d$, see  Fig.~\ref{fig:R_vs_freq}a.
The shape of the PEC patch can be arbitrary as long as the patches in any two cells are not connected (\ie the structure is low-pass).



The electric polarizability of a structure is a dyadic tensor $\vec{\gamma}$, which characterizes the ability of the structure to separate electrical charges. Consider the external uniform electrostatic field $\vec{E}_0$, applied to the structure. The polarizability relates this field to the induced electric dipole moment $\vec{p}$ in a unit cell of the structure~\cite{sjoberg2008}
\begin{equation}
	\label{eq:p_gamma}
    \vec{p} = \epsilon_0 \vec{\gamma}\vec{E_0}.
\end{equation}
Given the full electrostatic field $\vec{E}$ in this configuration (can be obtained by solving the electrostatic problem), we find the dipole moment $\vec{p}$ per unit cell ${\mathbb U}$ as
\begin{equation}
	\label{eq:p}
    	\vec{p}=\int_{\mathbb{U}}(\vec{\epsilon}-\epsilon_0)\vec{E}\mathrm{d}V+
    	\oint_{\partial\Omega}\vec{x}\hat{\vec{n}}\cdot(\vec{\epsilon}\vec{E})\mathrm{d}S.
\end{equation}
Here $\partial\Omega$ is the surface of the PEC subregion $\Omega$ within $\mathbb{U}$, $\epsilon$ is a dielectric permittivity tensor.
For example, consider Fig.~\ref{fig:R_vs_freq}a, where the subregion $\Omega$ is a PEC cross, and the substrate is dielectric.
To evaluate a component of the electric polarizability tensor $\gamma_{uv}$, $u,v\in\{x,y,z\}$, a $u$-component of the dipole moment $p_u$ due to $\vec{E}_0$ in $v$-direction should be considered.
Note that for the here considered structures $\gamma_{xz},\gamma_{yz},\gamma_{zx},\gamma_{zy}$ are negligible and $\gamma_{zz}$ is equal to polarizability of the substrate only.
Thus, we focus on evaluation of $\gamma_{uv}$, $u,v\in\{x,y\}$.
We consider here only the $xx$-component of the polarizability tensor. The other components are obtained in a similar manner, as discussed further in this paper.
In the rest of the paper we discuss the normalized quantity $\gamma_{xx}/A$, where $A=P_x\times P_y$.

There are existing polarizability estimation methods found in literature. 
\emph{The method of moments-based approach}~\cite{harrington1967} solves the first-order integral equation for potentials on $\partial \Omega$ with respect to the surface charge $\hat{\vec{n}}\cdot(\vec{\epsilon} \vec{E})$, which allows to evaluate~\eqref{eq:p}.
\emph{Variational approach}~\cite{sjoberg2008} provides a lower and an upper bounds on the electric polarizability as functionals of magnetic vector $\vec{F}$ and electric scalar $\phi$ potentials respectively for a given geometry with an applied uniform electric field.
\emph{Low-frequency asymptote approach} estimates the polarizability from a known reflection coefficient $R_{xx}$ of an $x$-polarized normally impinging wave at low frequency according to the expansion $R_{xx} = -\mathrm{j}k\gamma_{xx}/2A + \mathcal{O}(k^2)$ as $k\to 0$, which is valid  for a non-magnetic low pass periodic structures, not supporting electric currents in the longitudinal direction~\cite{sjoberg2009}.
Here, $k$ is the wavenumber in vacuum.

\section{Multimodal Network Approach}

The here \emph{proposed} method to determine electric polarizability is given in Sec.~IV.
It is based on the low-frequency asymptote approach and on the Multimodal Network Approach (MNA)~\cite{berral2015}.
In this section we shortly review MNA, as it will be beneficial for understanding of the proposed method.

MNA was developed for a frequency analysis of a plane wave scattering against periodic structures.
Consider an infinite structure, a fragment of which is depicted in Fig.~\ref{fig:R_vs_freq}a.
As we evaluate $\gamma_{xx}$ here, we will focus on expressing the corresponding reflection coefficient $R_{xx}$ of the structure using MNA.
Hence we consider a plane wave with polarization parallel to $x$-axis and the wavevector $\hat{\vec{k}}$ normal to the structure. 
The profile of the induced current density is $\vec{J}(x,y)=J_x(x,y)\hat{\vec{x}}+J_y(x,y)\hat{\vec{y}}$.
MNA suggests an equivalent transmission line scheme for this scattering configuration shown in Fig.~\ref{fig:R_vs_freq}b: $Y_{\mathrm{eq}}$ represents the PEC patch sheet, $Y^{\mathrm{(L),TE}}_{00}$ and $Y^{\mathrm{(R),TE}}_{00}$ are the admittances of the fundamental mode in the media on the left and on the right of the PEC patches respectively.
The base of MNA is a Floquet modes expansion of electric and magnetic fields within the structure.
Inherently $Y_\mathrm{eq}$ is a function of modal admittances, and given as~\cite{berral2015}
\begin{equation}
	\label{eq:Y_eq}
    Y_{\mathrm{eq}}(k) = \left[ \sum_{\Pi={\rm TE,TM}} \sum\nolimits'
    		 \dfrac{F^{\Pi}(k_{xn},k_{ym})}{Y_{mn}^{\mathrm{(L)},\Pi}(k) + Y_{mn}^{\mathrm{(R)},\Pi}(k)} \right]^{-1}
\end{equation}
\begin{equation}
	F^{\mathrm{TE}}(k_{xn},k_{ym}) = 
    \left|\dfrac{\tilde{J}_x(k_{xn},k_{ym})}{\tilde{J}_x(k_{x0},k_{y0})}\right|^2
    \dfrac{k_{ym}^2}{k_{xn}^2+k_{ym}^2},
\end{equation}
\begin{equation}
	F^{\mathrm{TM}}(k_{xn},k_{ym}) = 
    F^{\mathrm{TE}}(k_{xn},k_{ym})k_{xn}^2/k_{ym}^2,
\end{equation}
where $\tilde{J}_x(k_{xn},k_{ym})$ is the Fourier transform of $J_x(x,y)$, $k_{xn}=2\pi n/P_x$ and $k_{ym}=2\pi m/P_y$ are modal wavenumbers.
$Y^{\mathrm{(L/R)},\Pi}_{mn}$ are the total modal admittances, where the upper index stands for the media on the left (L) or the right (R) of the PEC structure and the mode's polarization $\Pi$ (TM or TE).
The sum $\sum '$ denotes $\sum_{(m,n)=-\infty}^\infty$ with the term $(m,n)=(0,0)$ excluded.
The equivalent network yields that the reflection coefficient of the fundamental mode is
\begin{equation}
\label{eq:R}
	R_{xx}(k) = \frac{Y_{00}^{(\mathrm{L}),\Pi}(k) - Y_{00}^{(\mathrm{R}),\Pi}(k)-Y_{\rm eq}}   {Y_{00}^{(\mathrm{L}),\Pi}(k) + Y_{00}^{(\mathrm{R}),\Pi}(k)+Y_{\rm eq}}.
\end{equation}

A validation and further details of MNA can be found in~\cite{berral2015,mesa2014}.
It is computationally effective and provides a deep insight in the physics of the problem.
However, a current profile $\vec{J}(x,y)$ is known in analytical form only for a few patch shapes, \eg a rectangle~\cite{berral2015} and a ring\cite{dubrovka2006}. For other shapes, obtaining a current profile $\vec{J}(x,y)$ and its Fourier transform requires additional treatment.
It was reported in~\cite{mesa2014} that $\vec{J}(x,y)$ can be extracted from a full-wave simulation on a single frequency point.

We note here that for computing $\gamma_{uv}$ one should consider an impinging wave with polarization in $v$-direction, and the corresponding $R_{uv}$.

\section{Proposed Method}

To introduce the \emph{proposed} method, we start with a treatment of a non-dielectric case (equivalently, $d=0$).
The dielectric layer is introduced at the end.
To use~\eqref{eq:Y_eq} we need to define the modal admittances.
For a material with a relative permittivity $\epsilon$ they are $Y_{mn,\epsilon}^{\rm TM}(k)=\epsilon k/(\eta_0\: \sqrt[]{\epsilon k^2-k_{xn}^2-k_{ym}^2})$  for TM harmonics and 
$Y_{mn,\epsilon}^{\rm TE}(k)=\sqrt[]{\epsilon k^2-k_{xn}^2-k_{ym}^2}/(k\eta_0)$  for TE,
where $\eta_0$ stands for the free space wave impedance.
In the non-substrate case we have free space ($\epsilon=1$) at the both sides of the structure.
Substitution of the modal admittances at~\eqref{eq:Y_eq} and~\eqref{eq:R} yields 
\begin{equation}
\begin{split}
	&R_{xx}(k) =\\ &-\left[ 1+\sum\nolimits'
    	\left|\dfrac{\tilde{J}_x(k_{xn},k_{ym})}{\tilde{J}_x(k_{x0},k_{y0})}\right|^2
        \dfrac{k^2-k_{xn}^2}{k\: \sqrt[]{k^2-k_{xn}^2-k_{ym}^2}}\right]^{-1}\! .
\end{split}
\end{equation}
For $k\to 0$ we have
\begin{equation}\nonumber
\begin{split}
	&R_{xx}(k)= \\&-\mathrm{j}k\left[ \sum\nolimits'
    	\left|\dfrac{\tilde{J}_x(k_{xn},k_{ym})}{\tilde{J}_x(k_{x0},k_{y0})}\right|^2
        \dfrac{k_{xn}^2}{\sqrt[]{k_{xn}^2+k_{ym}^2}}
        \right]^{-1} + \mathcal{O}(k^2).
\end{split}
\end{equation}
By comparing this expansion with the expansion in the base of the low-frequency asymptote approach,
we identify the electric polarizability per unit area in terms of the infinite sum
\begin{equation}
\label{eq:gamma_no_slab}
	\dfrac{\gamma_{xx}}{2A} = \left[ \sum\nolimits'
    	\left|\dfrac{\tilde{J}_x(k_{xn},k_{ym})}{\tilde{J}_x(k_{x0},k_{y0})}\right|^2
        \dfrac{k_{xn}^2}{\sqrt[]{k_{xn}^2+k_{ym}^2}}
        \right]^{-1}.
\end{equation}
This is the key formula of the proposed method.
Note that for a symmetric unit cell, the summation is over discrete points of an even function of both $k_{xn}$ and $k_{ym}$, which reduces the number of terms with approximately a factor of 4. For a non-symmetric unit cell, the number of terms is reduced approximately twice instead, since the function under the sum is symmetric against the line $k_{xn}=k_{ym}$ due to reality of $J_x(x,y)$.
The numerical implementation requires a truncation of the sum, and in the results presented in this paper $8\times 8$ terms were used.
The performance of~\eqref{eq:gamma_no_slab} is validated in Sec. V.

Let us generalize our result to include a dielectric substrate.
For this derivation we need to alter the modal admittances on the right side of $Y_{\mathrm{eq}}$.
See Fig.~\ref{fig:R_vs_freq}b, a dielectric slab of thickness $d$ backing up the patch array can be represented as a transmission line of length $d$ and corresponding admittance $Y_{mn,\epsilon}$, connected in series with the infinite transmission line with admittance $Y_{mn,1}=Y_{mn,\epsilon=1}$ representing free space.
The resulting admittance is
\begin{equation}
\label{eq:Y_R}
	Y_{mn}^{{\rm (R)},\Pi} = Y_{mn,\epsilon}^{\Pi}\dfrac{Y_{mn,1}^{\Pi}+\mathrm{j}\tan\left(d\:\sqrt[]{\epsilon k^2-k_{xn}^2-k_{yn}^2}\right)Y_{mn,\epsilon}^{\Pi}}{Y_{mn,\epsilon}^{\Pi}+\mathrm{j}\tan\left(d\:\sqrt[]{\epsilon k^2-k_{xn}^2-k_{yn}^2}\right)Y_{mn,1}^{\Pi}}.
\end{equation}
Performing the very same steps as done in the non-substrate case (substituting the modal admittances in~\eqref{eq:Y_eq} and~\eqref{eq:R} and taking a low frequency expansion) we end up with
\begin{equation}
	\label{eq:gamma_slab}
	\dfrac{\gamma_{xx}}{2A} = \left[\sum\nolimits' \left|\dfrac{\tilde{J}_x(k_{xn},k_{ym})}{\tilde{J}_x(k_{x0},k_{y0})}\right|^2
        \frac{k_{xn}^2M_{mn}}{\sqrt[]{k_{xn}^2+k_{ym}^2}}
        \right]^{-1} + \dfrac{(\epsilon-1)d}{2},
\end{equation}
\begin{equation}
\label{eq:Mmn}
	M_{mn}=
        \frac{2\left(\epsilon+\tanh\left[ d\: \sqrt[]{k_{xn}^2+k_{ym}^2} \right] \right)}{2\epsilon+(1+\epsilon^2)\tanh\left[ d\: \sqrt[]{k_{xn}^2+k_{ym}^2} \right]}.
\end{equation}
Note that $\epsilon=1$ or $d=0$ yields \eqref{eq:gamma_no_slab}.
The expression \eqref{eq:gamma_slab} is a good illustration of the physical insight of the proposed method.
The second term can be identified as a contribution of the  free-standing dielectric substrate.
The first term gives the contribution of the patch surface to the polarizability.
In each term we identify the same multiple as in the double sum of~\eqref{eq:gamma_no_slab}. 
The factor $M_{mn}$ accounts for how the presence of the dielectric enhances the polarizability of the patch array itself.
The validation of~\eqref{eq:gamma_slab} is shown in Sec.~V.

\section{Results and applications}

\begin{figure}[htbp]
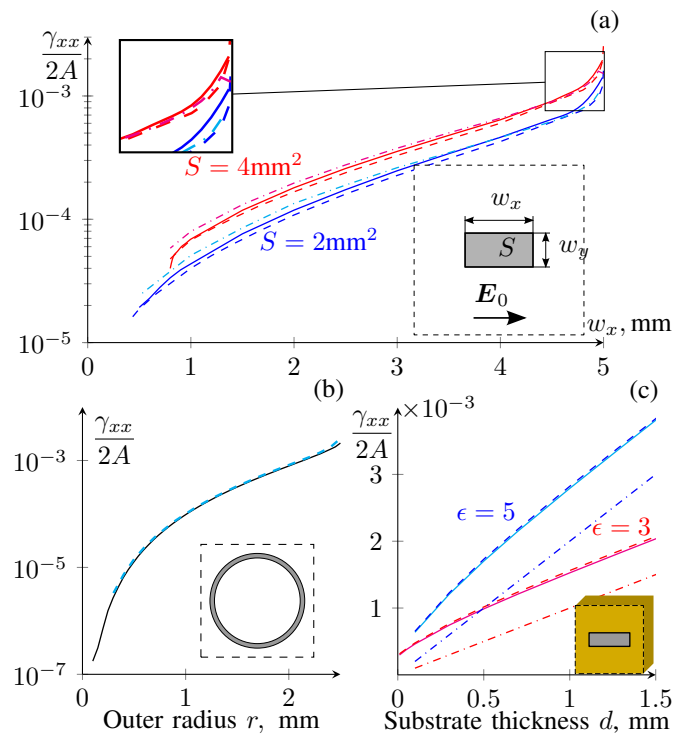

  \begin{minipage}{0.5\textwidth}
  \vspace{5pt}
  \centering
  \begingroup
	\tikzset{every picture/.style={scale=1}}
	\input{const_A.tikz}
  \endgroup
    
  \vspace{-5pt}

  \end{minipage}
    \begin{minipage}{0.24\textwidth}
  \vspace{-0pt}
  \centering
  \begingroup
	\tikzset{every picture/.style={scale=1}}
%
%
\definecolor{mycolor1}{rgb}{0.00000,0.44700,0.74100}%
\definecolor{mycolor2}{rgb}{0.85000,0.32500,0.09800}%
\definecolor{mycolor3}{rgb}{0.92900,0.69400,0.12500}%
\begin{tikzpicture}[
spy using outlines={red, magnification=3, size=25 * 1.5,
                          connect spies}]

\begin{axis}[%
axis lines = middle,
/pgfplots/axis x line=bottom,
width=1.35in,
height=1.4in,
at={(0.758in,0.481in)},
scale only axis,
xmin=0,
xmax=2.5,
xlabel style={font=\color{white!15!black},at={(axis description cs:0.5,-0.1)},anchor=north},
xlabel={Outer radius $r,\text{ mm}$},
ymode=log,
ymin=9.9999e-08,
ymax=0.01,
yminorticks=true,
ylabel style={font=\color{white!15!black}},
ylabel={$\dfrac{\gamma_{xx}}{2A}$},
axis background/.style={fill=white},
legend style={at={(0.03,0.97)}, anchor=north west, legend cell align=left, align=left, draw=white!15!black}
]
\addplot [color=black,line width=0.5pt]
  table[row sep=crcr]{%
1.000000e-01   1.765515e-07 \\ 
1.489796e-01   3.124249e-07 \\ 
1.979592e-01   7.106387e-07 \\ 
2.469388e-01   1.544108e-06 \\ 
2.959184e-01   2.667069e-06 \\ 
3.448980e-01   4.203421e-06 \\ 
3.938776e-01   6.368038e-06 \\ 
4.428571e-01   9.008267e-06 \\ 
4.918367e-01   1.219026e-05 \\ 
5.408163e-01   1.624896e-05 \\ 
5.897959e-01   2.093028e-05 \\ 
6.387755e-01   2.628522e-05 \\ 
6.877551e-01   3.274749e-05 \\ 
7.367347e-01   4.000309e-05 \\ 
7.857143e-01   4.804719e-05 \\ 
8.346939e-01   5.741590e-05 \\ 
8.836735e-01   6.781166e-05 \\ 
9.326531e-01   7.904232e-05 \\ 
9.816327e-01   9.188947e-05 \\ 
1.030612e+00   1.059644e-04 \\ 
1.079592e+00   1.209882e-04 \\ 
1.128571e+00   1.378538e-04 \\ 
1.177551e+00   1.562849e-04 \\ 
1.226531e+00   1.757065e-04 \\ 
1.275510e+00   1.972626e-04 \\ 
1.324490e+00   2.207858e-04 \\ 
1.373469e+00   2.453399e-04 \\ 
1.422449e+00   2.723815e-04 \\ 
1.471429e+00   3.018588e-04 \\ 
1.520408e+00   3.325186e-04 \\ 
1.569388e+00   3.659338e-04 \\ 
1.618367e+00   4.026041e-04 \\ 
1.667347e+00   4.404863e-04 \\ 
1.716327e+00   4.816173e-04 \\ 
1.765306e+00   5.270000e-04 \\ 
1.814286e+00   5.738448e-04 \\ 
1.863265e+00   6.244394e-04 \\ 
1.912245e+00   6.809157e-04 \\ 
1.961224e+00   7.393248e-04 \\ 
2.010204e+00   8.020873e-04 \\ 
2.059184e+00   8.737244e-04 \\ 
2.108163e+00   9.479914e-04 \\ 
2.157143e+00   1.028044e-03 \\ 
2.206122e+00   1.122424e-03 \\ 
2.255102e+00   1.222091e-03 \\ 
2.304082e+00   1.330175e-03 \\ 
2.353061e+00   1.468449e-03 \\ 
2.402041e+00   1.621312e-03 \\ 
2.451020e+00   1.798634e-03 \\ 
2.500000e+00   2.117805e-03 \\ 
};
\addlegendentry{Proposed}

\addplot [color=cyan,dashed,line width=0.9pt]
  table[row sep=crcr]{%
3.000000e-01   3.311306e-06 \\ 
4.000000e-01   7.401085e-06 \\ 
5.000000e-01   1.424665e-05 \\ 
6.000000e-01   2.371749e-05 \\ 
7.000000e-01   3.659447e-05 \\ 
8.000000e-01   5.288791e-05 \\ 
9.000000e-01   7.475153e-05 \\ 
1   1.002515e-04 \\ 
1.100000e+00   1.336347e-04 \\ 
1.200000e+00   1.708085e-04 \\ 
1.300000e+00   2.158948e-04 \\ 
1.400000e+00   2.684098e-04 \\ 
1.500000e+00   3.318887e-04 \\ 
1.600000e+00   4.010135e-04 \\ 
1.700000e+00   4.878031e-04 \\ 
1.800000e+00   5.838252e-04 \\ 
1.900000e+00   6.941655e-04 \\ 
2   8.235766e-04 \\ 
2.100000e+00   9.783266e-04 \\ 
2.200000e+00   1.158478e-03 \\ 
2.300000e+00   1.401809e-03 \\ 
2.400000e+00   1.754804e-03 \\ 
2.490000e+00   2.476116e-03 \\ 
};
\addlegendentry{CST}


\legend{};
\end{axis}



\filldraw[fill=white, draw=black, dashed] (3.5,1.45) rectangle ++(1.5,1.5);
\filldraw[fill=black!40!white, draw=black, line width=0.05pt] (4.25,2.2) circle (0.63);
\filldraw[fill=white, draw=black, line width=0.05pt] (4.25,2.2) circle (0.57);
	\node at (5.2,5.0) {(b)};
\end{tikzpicture}%
  \endgroup
  
  \vspace{-5pt}

  \end{minipage}
  \begin{minipage}{0.24\textwidth}
  \vspace{-0pt}
  \centering
  \begingroup
	\tikzset{every picture/.style={scale=1}}
	\input{slab.tikz}
  \endgroup
    
  \vspace{-5pt}

    \end{minipage}
  \caption{(a)~Normalized polarizability of an array of rectangular patches in free space with a fixed surface area $S=w_x\times w_y$.
    (b)~Normalized polarizability of an array of ring-shaped patches.
  (c)~Normalized polarizability of an array of 3mm$\times$1mm rectangular patches with a dielectric substrate of thickness $d$ and $\epsilon=\{ 3, 5 \}$.
  Details and discussion of the figures is in Sec. V.
}
  \label{fig:gamma_ratio}
\end{figure}

We consider a rectangular PEC patch array in free space with the unit cell size $P_x \times P_y = 5$mm$\times 5$mm, the total area of the patch taking two values $S=w_x\times w_y=\{2,4\}$mm$^2$.
In the proposed method, the analytic current profile~\cite{berral2015} was used.
In Fig.~\ref{fig:gamma_ratio}a the normalized polarizability $\gamma_{xx}/2A$, calculated with our proposed method (solid curves) from~\eqref{eq:gamma_no_slab}, by the upper bound of variational approach~\cite{sjoberg2009} (dash-dotted curves) computed in COMSOL and by the low-frequency asymptote approach based on simulations in CST MW Studio (dashed curves), is shown as a function of the patch length $w_x$.
The curves agree well, and also the behavior as $w_x\to P_x$ is captured by all the approaches.
At the limit point $w_x=P_x$ the structure becomes an array of infinitely long stripes and cease to be low-pass in the $x$-direction.
This explains a higher increase rate of the polarizability in this region.
One can also observe a deviation between the proposed method and the variational approach when the patch is short and wide ($w_x\sim 1$mm), and also when $w_x$ approaches $P_x$. 
This is mainly due to a limited validity of the analytic current profile $\vec{J(x,y)}$ in these extreme cases. The polarizabilities of an array of ring-shaped patches ($P_x \times P_y = 5$mm$\times 5$mm, ring width 0.1mm, analytic current profile~\cite{dubrovka2006}) are shown in Fig.~\ref{fig:gamma_ratio}b, calculated with our proposed method (solid curves), and the low-frequency asymptote, based on CST results (dashed curves). Good agreement between the two methods is observed.

A physical limitation on the bandwidth of transmission blockage of electromagnetic waves by a periodic structure is given by~\cite{gustafsson2009}
\begin{equation}
	(\lambda_2-\lambda_1)\ln\dfrac{1}{T_0}\leq \pi^2\dfrac{\gamma}{2A},
    \label{eq:BW_sumrule}
\end{equation}
where the wavelength range between $\lambda_1$ and $\lambda_2$ has a transmission coefficient magnitude no higher than some fixed value $T_0$.
Thus, the total attainable bandwidth is limited from above by polarizability.
We observe in Fig.~\ref{fig:gamma_ratio}a that with the same amount of material we can change the value of polarizability by 100 times just by altering the shape of the patch.
Thus choosing a proper patch shape, i.e. the support for the current profile function $\vec{J(x,y)}$, can drastically increase the operational bandwidth of a frequency selective structure.

To illustrate the application of the proposed approach in a bandwidth optimization, we consider an example of the blockage bandwidth for a low-pass structure consisting of rectangles. The goal here is to find the smallest rectangular area $S=w_x\times w_y$ reaching a blockage bandwidth $B_0$, given the periodicity $P_x,P_y$ and constraints $w_x\leq a$, $w_y\leq b$. To be specific, assume $P_x=P_y=5$mm, $w_x\leq 2$mm, $w_y\leq 4$mm and the bandwidth $B_0$ corresponds according to~\eqref{eq:BW_sumrule} to the polarizability $\gamma=10^{-4}$. By inspection of a parametric sweep in Fig.~\ref{fig:gamma_ratio}a, we see, that $w_x=2$mm, $w_y=1$mm with surface area $S=2$mm$^2$ minimizes such a problem. 

For other shapes of a patch element, we can set up an efficient parametrization for optimization procedures with e.g. genetic algorithms. Assume we have obtained the current profile $J_x(x,y)$ for a given patch element using e.g. a full-wave solver~\cite{mesa2014}. For the family of shapes, obtained from the original shape under local stretching and scaling of the shape’s parts, the obtained $J_x(x,y)$, modified with the corresponding local stretching and scaling, is expected to remain a good approximation for a current profile, as indicated by numerical examinations. Bandwidth optimization would thus be applied to parameter variations where the appropriately modified $J_x(x,y)$ is still valid. Observe that such a local rescaling of $J_x(x,y)$ and a calculation of truncated sums in the proposed method would be significantly faster than the solution of an electrostatic problem for each parametric sweep point.

We also verified our method for structures with dielectric slab inclusions.
Consider an infinite array of rectangular patches backed up by a dielectric substrate.
The patch size is fixed to 3mm$\times 1$mm and the relative permittivity of the substrate takes two values $\epsilon=\{3,5\}$.
Fig.~\ref{fig:gamma_ratio}c depicts the normalized polarizability as a function of the substrate thickness $d$.
The solid curves correspond to the proposed method, dashed represent the results of a low frequency asymptotic extraction from full-wave simulation (CST).
The dot-dashed lines are polarizabilities of a free-standing dielectric slab of the corresponding material and thickness.
We observe that there is a good agreement between the result of~\eqref{eq:gamma_slab} and the CST-based estimates.
We can see that the total electric polarizability indeed has a linear term representing the dielectric substrate contribution.
The difference between the solid and dot-dashed curves (that is the patch array contribution) seems to be approximately constant as $d$ becomes greater than some value around 0.8mm.
Analysis of the second factor in~\eqref{eq:Mmn} for the lowest order term (e.g. $(n,m)=(1,0)$) shows that the factor is almost independent of $d$ when $d\geq P_x/(2\pi)\approx 0.8$mm.
Subsequently the total contribution in the polarizabilty from the metal patches is approximately constant as $d>0.8$mm. 

\section{Conclusion}
The here proposed method to evaluate electric polarizability has shown a fine agreement with the two alternative approaches.
The proposed method allows to set up a numerically efficient sweep of geometrical parameters according to the strategy introduced in Sec.~V.
Such a parametric sweep, combined with the sum rules relating polarizability and operational bandwidth (see \eg~\cite{gustafsson2011,gustafsson2009}), can be used in a bandwidth optimization.
Additionally, the proposed method provides an intuitive understanding of which parts of a structure produce the most significant contributions to the polarizability value, see a discussion in the end of Sec.~V for an example of a patch array on a dielectric substrate.

The main limitation of the proposed method arise in finding a good enough approximation of the electric current's profile for the patches of arbitrary shape.
Techniques to overcome this limitation is one of the further research directions on this topic and it closely related with Multimodal Network Approach.
Another interesting directions are to implement several layers of PEC structures~\cite{torres2016} and PEC layers of finite non-negligible thickness.
\section*{Acknowledgment}
We gratefully  acknowledge  the  support  of  the SSF grant AM13-0011 and VINNOVA project ChaseOn/IAA.
We are also grateful for discussions with Prof. Francisco Mesa and Prof. Raul Rodr{\'i}guez-Berral.

\ifCLASSOPTIONcaptionsoff
  \newpage
\fi

\end{document}